\documentclass[aps,prc,preprint,groupedaddress]{revtex4}

\newcommand{\beq}{\begin{equation}}
\newcommand{\eeq}{\end{equation}}
\newcommand{\beqa}{\begin{eqnarray}}
\newcommand{\eeqa}{\end{eqnarray}}

\usepackage{epsfig}
\usepackage{youngtab}

\begin{document}

\title{The role of five-quark components 
in gamma decay of the $\Delta(1232)$}

\author{Q. B. Li}
\email[]{ligb@pcu.helsinki.fi}
\affiliation{Helsinki Institute of Physics,
POB 64, 00014 University of Helsinki, Finland}

\author{D. O. Riska}
\email[]{riska@pcu.helsinki.fi}
\affiliation{Helsinki Institute of Physics, POB 64,
00014 University of Helsinki, Finland}

\thispagestyle{empty}

\date{\today}
\begin{abstract}
An admixture of 10 - 20\% of $qqqq\bar q$ components in 
the $\Delta(1232)$ resonance is shown to
reduce the well known underprediction by the $qqq$ quark
model of the decay width for
$\Delta(1232)\rightarrow N\gamma$ decay by about half
and that of the corresponding
helicity amplitudes from a factor $\sim$ 1.7 to
$\sim$ 1.5.
The main effect is due to the quark-antiquark
annihilation transitions: $qqqq\bar q\rightarrow qqq\gamma$, 
the consideration of
which brings the ratio $A_{3\over 2}/A_{1\over 2}$ 
and consequently the $E2/M1$ ratio $R_{EM}$ into agreement
with the empirical value.  Transitions between the $qqqq\bar q$ 
components in the resonance and the nucleon: 
$qqqq\bar q \rightarrow qqqq\bar q\gamma$,
are shown to enhance the
calculated decay width by only a few percent, as long as the
probability of the
$qqqq\bar q$ component of the proton and the $\Delta(1232)$ 
is at most 20\%.
The transitions $qqqq\bar q \rightarrow qqqq\bar q\gamma$
between the $qqqq\bar q$ components in
the $\Delta(1232)$
and the proton do not lead to a nonzero value
for $R_{EM}$. 

\end{abstract}

\pacs{}

\maketitle

\section{Introduction}

The 3 valence quark model for the baryons does not provide a quantitatively
satisfactory description of the
electromagnetic and strong decay widths of the lowest energy 
nucleon resonances. Even in refined versions of that model the 
calculated decay 
width of the $\Delta(1232)$ is typically less than half of the empirical
value, while those of the $N(1440)$ and the $\Delta(1600)$
are even smaller \cite{bruno, melde}. Explicit coupled 
channel treatments of the interacting $\pi-N-\Delta$ system
\cite{sato, sato2} show that this problem is likely to
arise from coupling to
explicit pion degrees of freedom, which are missing in
the $qqq$ valence
quark model. This suggests that the quark model be extended
to include explicit sea-quark ($q\bar q$) configurations. 
The presence of such sea-quark configurations in the proton
has in fact been experimentally demonstrated 
\cite{towell,NA51,NMC,HERMES}.

Here such an extension of the valence quark model is made to include
those $qqqq\bar q$ components in the proton and the
$\Delta(1232)$ wave function, which are likely to require
the lowest excitation energy, in order
to study the effect of these $qqqq\bar q$ components on
the calculated 
electromagnetic decay rates of the $\Delta(1232)$.
As in a previous study of the effect 
of $qqqq\bar q$ components on the calculated pion decay
rate of the $\Delta(1232)$, it is found that transitions
between the $qqqq\bar q$ components in the resonances
and in the nucleon themselves leads to modifications of only a 
few percent \cite{delpion}. On the other hand the 
direct quark-antiquark annihilation transitions between
the $qqqq\bar q$ components in the resonances and the
$qqq$ component of the proton is significant. 

The confining and hyperfine interactions
between the quarks can also trigger such quark-antiquark 
annihilation transitions. In the case of pion decay 
the confining interaction 
leads to an enhancement of the net effect of annihilation 
transitions on the calculated decay rates. Here it is found
to have the
opposite effect in the case of the electromagnetic decays,
unless the interaction potential is attractive at short range .
The magnitude of the confinement triggered annihilation
transitions is estimated with two schematic models
(linear and harmonic) for the confining interaction.
Given the opposite effects of the confinement triggered
annihilation transitions in pion and electromagnetic
decay, the conclusion is that this effect should be
of minor net significance.

It is found that the effect of the direct annihilation transitions 
on the calculated helicity amplitudes for gamma decay
of $\Delta(1232)$ is to bring them closer to the empirical
values. A probability of $\sim$ 20\% for the $qqqq\bar q$
component in the $\Delta(1232)$ leads to a reduction of the
underprediction of $\sim 1.7$ in the valence quark model
to $\sim 1.5$. This leads to a corresponding
reduction of the underprediction of the radiative width by
about half.

The quark-antiquark annihilation transitions bring the 
calculated helicity amplitude ratio $A_{3/2}/A_{1/2}$ into
agreement with the empirical ratio. As a consequence they
lead to a non-vanishing value for the calculated $E2/M1$ 
ratio $R_{EM}$, which falls within the empirical
range.

The paper is structured in the following way.
In section II 
the effect of transitions between the $qqqq\bar q$ components 
wave functions of proton and $\Delta(1232)$ in $qqqq\bar q$ 
configurations
are calculated. 
In Section III the corresponding annihilation transitions
$qqqq\bar q\rightarrow qqq\gamma$ are considered.
Section IV contains a concluding
discussion. 

\section{Transitions between $qqqq\bar q$ components 
in $\Delta^+\rightarrow p\gamma$ decay}

Consider $\Delta^+\rightarrow p\gamma$ decay that arises
from the direct electromagnetic coupling to 
constituent quarks: $qq\gamma$. 
To lowest order in the photon momentum the transition
amplitude obtained from the electromagnetic $\gamma_\mu$
coupling for pointlike quarks is then 
\beqa
T_i=\frac{e_i}{2m}{{\sigma_i}_-} \sqrt{k_\gamma},
\label{dt}
\eeqa
where $e_i$ and $m$ are the electric charge and mass of the 
quark that emits the photon, respectively. The momentum
of the final right-handed photon is taken to be in the direction of the
$z$-axis:
$\vec k =(0,0,k_\gamma)$ with $k_\gamma$=259 MeV in the center of 
mass frame of $\Delta(1232)$.

The $\Delta^+\rightarrow p\gamma$ transition is described 
by the two 
independent helicity amplitudes:
\beqa
&& A_{\frac{3}{2}}=\langle p,\frac{1}{2}~\frac{1}{2}~| 
~\sum_i^{n_q}~T_i~ |~\Delta^+,~\frac{3}{2}~\frac{3}{2}\rangle\,,
\nonumber\\
&& A_{\frac{1}{2}}=\langle p,~\frac{1}{2}~ -\frac{1}{2}~| 
~\sum_i^{n_q} T_i~ |~\Delta^+,~\frac{3}{2}~\frac{1}{2}\rangle~.
\label{helam}
\eeqa
These represent the helicity components $3/2$ 
and $1/2$ of the $\Delta(1232)$ on the direction of the
photon momentum.  
In eq. (\ref{helam}) $n_q$ is the 
number of constituent quarks.

The spatial wave function of the quarks in the spatially
symmetric ground state will be schematically
described by the harmonic oscillator wave function:
\begin{equation}
\varphi_0(\xi_i) = ({\omega_3^2\over\pi})^{3/4}\,
e^{-{\xi}_i^2\,\omega_3^2/2}\, .
\label{harmosc}
\end{equation}
The scale of the oscillator parameter $\omega_3$ 
may be set by 
the empirical radius of the
proton as $\omega_3 = 1/r_p \simeq 225$ MeV. In (\ref{harmosc})
${\vec \xi}_i$, 
i=1, 2, are the 
standard Jacobi coordinates of the 3-quark system.
The  helicity amplitudes for $\Delta^+\rightarrow p\gamma$ decay in 
the conventional $qqq$ configuration are then:
\beqa
&& A_{\frac{3}{2}}^{(3q)}=-\frac{\sqrt{6}}{3}\frac{e}{2m} \sqrt{k_\gamma}
~(1-{k_\gamma\,^2\over 6 \omega_3^2})\, , \nonumber\\
&& A_{\frac{1}{2}}^{(3q)}=-\frac{\sqrt{2}}{3}\frac{e}{2m} \sqrt{k_\gamma}
~(1-{k_\gamma\,^2\over 6 \omega_3^2}).
\label{hel3q}
\eeqa
Here $m$ is again the constituent quark mass. These expressions yield the
usual quark model value for the ratio of the helicity amplitudes:
$A_{\frac{3}{2}}^{(3q)}/A_{\frac{1}{2}}^{(3q)}=\sqrt{3}$.
If the constituent quark mass is taken to be 340 MeV, these
expressions lead to the values $A_{\frac{1}{2}}^{(3q)}=
-0.083/\sqrt{\mathrm{GeV}}$ and $A_{\frac{3}{2}}^{(3q)}=
-0.143/\sqrt{\mathrm{GeV}}$. These values are smaller by factors
1.6 and 1.8,  respectively, than the corresponding
experimental values $ A_{\frac{1}{2}} =
-0.135\pm0.006/\sqrt{\mathrm{GeV}}$
and $ A_{\frac{3}{2}} = -0.255\pm0.008/\sqrt{\mathrm{GeV}}$
\cite{PDG}.

Consider now $qqqq\bar q$ admixtures in the proton and the
$\Delta^+$. Positive parity demands that these have to
be $P-$wave states. 
The spin dependence of the hyperfine interaction between the
quarks implies that the $qqqq\bar q$ configurations that
have the lowest energy, and which are most likely to
form notable admixtures in the baryon states, 
are those that have the most antisymmetric
$qqqq$ configurations, which are compatible with the requirement
of overall antisymmetry. In the case of the nucleon
this state has the mixed
spin-flavor symmetry 
$[4]_{FS}[22]_F[22]_S$
and in the case of the $\Delta(1232)$ the mixed
spin-flavor symmetry
$[4]_{FS}[31]_F[31]_S\,\,$ \cite{helminen}.  

The  $qqqq\bar q$ component in the proton, with a
$qqqq$ configuration with
spin-flavor symmetry $[4]_{FS}[22]_F[22]_S$  
has mixed spatial symmetry $[31]_X$, and may be
represented by the wave function: 
\beqa
&&\psi_p(M_S) ={A_{p5}\over \sqrt{2}}\sum_{a,b}
\sum_{m,s} (1,1/2,m,s\vert\, 1/2,M_S)\,
C^{[1111]}_{[211]a,[31]a}
\nonumber\\
&&[211]_C(a)\,[31]_{X,m}(a)\, [22]_F(b)\,[22]_{S}(b)\, \bar\chi_s\,
\phi(\{r_i\})\,.
\label{5qp}
\eeqa
Here $M_S$ denotes the spin-z component of the state and $A_{p5}$ 
is the amplitude of the configuration. The symbol
$C^{[1111]}_{[211]a,[31]a}$ is a  $S_4$ Clebsch-Gordan
coefficient.
The color, space and flavor-spin wave functions of the
$qqqq$ subsystem have here been denoted by their Young patterns
respectively. The sum over $a$ runs over the 3 
configurations of the $[211]_C$ and $[31]_X$ representations
of $S_4$, and the sum over $b$ runs over the 2 
configurations of the
$[22]$ representation of $S_4$ respectively \cite{chen}.  
Note that as the isospin of the $qqqq$ of the $[22]_F$
configuration is 0, the antiquark can only be a $\bar d$
quark in this configuration.

The corresponding $qqqq\bar q$ configuration in the
$\Delta(1232)$, with the flavor-spin symmetry 
$[4]_{FS}[31]_F[31]_S$ in 
the $qqqq$ configuration 
may be represented by the wave function:
\beqa
&&\psi_{\Delta^+}^{(J)}(M_S) ={A_{\Delta 5}^{(J)}\over \sqrt{3}}\sum_{a,b}
\sum_{m,s,M,j;T,t}
(1,1,m,M\vert\, J,j)(J,1/2,j,s\vert\, 3/2,M_S)\,
C^{[1111]}_{[211]a,[31]a}\,
\nonumber\\
&&(1,1/2,T,t|3/2,1/2)[211]_C(a)\,[31]_{X,m}(a)\, [31]_{F,T}(b)
\,[31]_{S,M}(b)\, \bar\chi_{t,s}\,
\phi(\{r_i\})\, .
\label{q5d}
\eeqa
Here $J$ denotes the total angular momentum of the
$qqqq$ system, which takes the values 1 and 2,
and $A_{\Delta 5}^{(J)}$ is the amplitude
of the configuration in the $\Delta(1232)$. 
The sum over $a$ again runs over the 3 
configurations of the $[211]_C$ and $[31]_X$ representations
of $S_4$. Here the sum over $b$ runs over the 3 
configurations of the $[31]$ representation.  
Here the isospin-z component of the 4-quark state is
denoted $T$ and that of the antiquark $t$, which results in 
the quark combination
$uudd\bar d$ for T=0, t=1/2 and $uuud\bar u$ for T=1, t=$-1/2$. 
Since there is no isospin flip 
in the transition operator (\ref{dt}), only the five quark 
configuration $uudd\bar d$ in 
$\Delta^+$ contributes to the direct transition 
$\Delta^+\rightarrow p\gamma$. 

The orbital wave function of the P-shell $qqqq$ states $[31]_X$ in 
eqs. (\ref{5qp}) and (\ref{q5d}) are described by the product of 
the S-wave and P-wave harmonic oscillator functions: 
\begin{equation}
\tilde\varphi_0(\xi_i) = ({\omega_5^2\over\pi})^{3/4}\, 
e^{-\xi_i^2\,\omega_5^2/2}\, ,~~~~~~~
\tilde\varphi_{1m}(\xi_i) = \sqrt{2}\omega_5\xi_{i,m}\varphi_0(\xi_i)\,.
\label{harmosc5}
\end{equation}
Here the oscillator parameter $\omega_5$ is that for the 
$qqqq \bar q$ system. The operators
$\xi_i$, i=1..3, are the
standard Jacobi coordinates for the five-quark system 
in the spherical
basis \cite{helminen}.     
   
The calculation of helicity amplitudes for the transition between
the $qqqq\bar q$ components in proton and $\Delta$ is straightforward
and leads to: 
\beqa
&& A_{\frac{3}{2}}^{(5q)}=
-\frac{2\sqrt{3}}{9}(\delta_{J1}+\sqrt{5}\delta_{J2}) 
\frac{e}{2m} \sqrt{k_\gamma} 
~(1-{k_\gamma\,^2\over 5 \omega_5^2})\, ,\nonumber\\ 
&& A_{\frac{1}{2}}^{(5q)}=-\frac{2}{9}
(\delta_{J1}+\sqrt{5}\delta_{J2}) \frac{e}{2m} \sqrt{k_\gamma} 
~(1-{k_\gamma\,^2\over 5 \omega_5^2})\, .
\label{hel5q}
\eeqa

From eqs. (\ref{hel3q}) and (\ref{hel5q}) one obtains the
ratio between the helicity 
amplitudes for direct transition when the $qqqq\bar q$ 
configurations in proton and $\Delta^+(1232)$ are included
to be:
\beqa
\frac{A_{\frac{3}{2}}}{A_{\frac{1}{2}}}=
\frac{A_{p3}A_{\Delta3}A_{\frac{3}{2}}^{3q}+
A_{p5}A_{\Delta 5}^{(J)}A_{\frac{3}{2}}^{5q}} 
{A_{p3}A_{\Delta3}A_{\frac{1}{2}}^{3q}+A_{p5}
A_{\Delta 5}^{(J)}A_{\frac{1}{2}}^{5q}}=\sqrt{3},
\label{r31}       
\eeqa
which is the standard quark model result.
Here $A_{p3}$, $A_{p5}$ are the amplitudes for the
$qqq$ and $qqqq\bar q$ components of the proton and  the
$\Delta(1232)$ respectively, and 
$A_{\Delta 3}$, $A_{\Delta 5}^{(J)}$ 
are the amplitudes for the corresponding components of the
$\Delta(1232)$. 

The magnetic dipole $M_1$ and electric quadrupole $E_2$ moment 
contributions to $\Delta^+\rightarrow p\gamma$ decay
are related to the helicity amplitudes as \cite{PDG}
\beqa
&& M_1=-\frac{1}{2\sqrt{3}}~
(3A_{\frac{3}{2}}+\sqrt{3}A_{\frac{1}{2}})\, ,\nonumber\\
&& E_2=\frac{1}{2\sqrt{3}}~
(A_{\frac{3}{2}}-\sqrt{3}A_{\frac{1}{2}}).
\label{m1e2}
\eeqa

Since the contribution from the direct transitions 
between $qqqq\bar q$ components of the $\Delta(1232)$ and
the nucleon
leaves the calculated 
ratio $A_{\frac{3}{2}}/A_{\frac{1}{2}}$ 
unchanged from the value $\sqrt{3}$ given by the $qqq$ configuration           
with spatially
symmetric wave functions(\ref{r31}), they leave the $E2$ amplitude
unchanged at 0.

While direct transitions between the $qqqq\bar q$ 
components in proton and $\Delta$ do not change the 
calculated value for the 
$E_2/M_1$ ratio for 
$\Delta^+\rightarrow p\gamma$ decay, they do affect the
calculated
decay width. 
In terms of the helicity amplitudes, the decay width 
is given by \cite{PDG}, 
\beqa
\Gamma=\frac{k^2_\gamma}{\pi}~\frac{2M_p}
{(2J+1)M_\Delta}[|A_{\frac{3}{2}}|^2+|A_{\frac{1}{2}}|^2],
\label{width}
\eeqa
\vspace{0.2cm}
where $M_p$ and $M_\Delta$ are the masses of proton and 
the $\Delta$, respectively, and $J=3/2$ is the 
spin of $\Delta$. By taking into account the normalization of the wave 
functions of proton and $\Delta(1232)$
in both the $qqq$ and $qqqq\bar q$ configurations, the enhancement 
of the calculated decay width that arises from the direct transitions
between the $qqqq\bar q$ components is:
\beqa
\delta=\frac{\Gamma}{\Gamma_{3q}}=
\frac{{\displaystyle{\sum_{\lambda=
1/2,3/2}} |A_{p3}A_{\Delta 3}A_\lambda^{(3q)}+
A_{p5}A_{\Delta 5}^{(J)} A_\lambda^{(5q)}|^2}}
{{\displaystyle|A_{\frac{3}{2}}^{(3q)}|^2+|A_{\frac{1}{2}}^{(5q)}|^2}}. 
\label{enhance}
\eeqa  
With the present wave function model, (\ref{hel3q}), (\ref{hel5q}), 
this leads to the expression:
\beqa
\delta=|A_{p3} A_{\Delta 3}+\frac{\sqrt{2}}{3}
A_{p5}(A_{\Delta 5}^{(1)}+\sqrt{5}A_{\Delta 5}^{(2)})|^2\, ,
\eeqa   
if the radii of the 3- and the 5-quark components 
are taken to be equal, so that
\begin{equation}
\omega_5 = \sqrt{{6\over 5}}\omega_3\, .
\label{omega}
\end{equation}
It is worth noting that the condition (\ref{omega}) is not
necessary for the eq. (\ref{r31}).
The numerical effect of the 5 quark components is small
even when the probability of the $qqqq\bar q$ components of
the proton and the $\Delta(1232)$ are larger than 10\%.
With 10\% probability for the $qqqq\bar q$ component
in the nucleon and $\Delta(1232)$, in which the 
proportion of the $J=1$ and $J=2$ $qqqq$ states in 
$\Delta(1232)$ are assumed 
to be 50\% and 50\%, respectively,
the effect of the $qqqq\bar q$ component is to enhance the calculated
values of both helicity
amplitudes by a factor $\sqrt{\delta} \sim 1.01$, and the
decay width by a factor $\delta \sim 1.02$. With the
enhancement factor 1.01 the helicity amplitude 
$A_{{1\over 2}}$ is smaller than the empirical value
by a factor 1.6, and the calculated value of
$A_{{3\over 2}}$ smaller by a factor 1.7 than the
corresponding empirical value. Hence the net effect of the 
transition between the $qqqq\bar q$ component in the $\Delta(1232)$
and proton is to increase the decay width 
by only a few percent at most. 

\section {$q\bar q$ annihilation transitions 
in $\Delta^+\rightarrow p\gamma$ decay}

\subsection{Direct annihilation transitions}

\begin{figure}[t]
\vspace{20pt} 
\begin{center}
\mbox{\epsfig{file=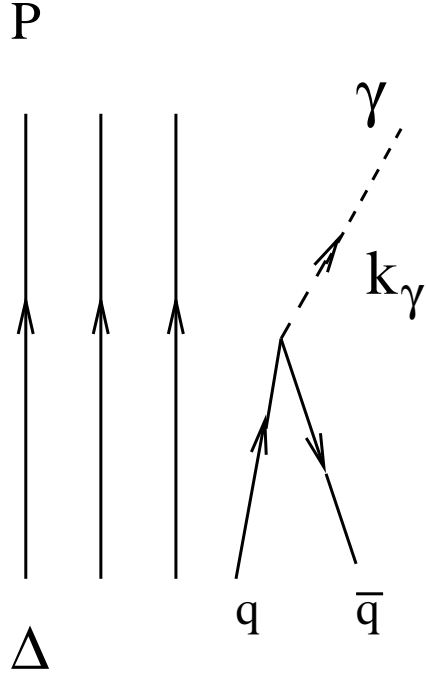, width=55mm}} 
\caption{Direct $qqqq\bar q\rightarrow qqq\gamma$ annihilation
process}
\label{fig1}
\end{center}
\vspace{10pt} 
\end{figure}

The Dirac ($\gamma_\mu$) coupling for pointlike quarks
used in previous section leads to following
the $q\bar q\rightarrow \gamma$
transition operator for the transition $\Delta^+\rightarrow p\gamma$ 
illustrated in
in Fig. \ref{fig1}:
\beqa
T_{a}=\sum_{i=1}^4 e_i {\sigma_i}_- \frac{1}{\sqrt{k_\gamma}},
\label{at}
\eeqa
where $e_i$ is the electric charge of the quark that annihilates
the 
anti-quark and 
${\sigma_i}_-$ is the spin lowering operator.
The ${\sigma_i}_-$
in eq. (\ref{at}) requires the annihilating quark to have the 
spin-z component $1/2$, which, in combination with the anti-quark 
with spin-z component $1/2$, produces the 
final photon with angular moment $L=1, L_z =1$. Note there
is no contribution from transitions of the reverse type
$qqq\rightarrow qqqq\bar q$.

The helicity amplitudes for the $qqqq\bar q\rightarrow qqq+\gamma$ 
transition 
are obtained as matrix elements of the operator (\ref{at}) between the 
proton in $qqq$ 
configuration and $\Delta$ in $qqqq\bar q$ configuration (\ref{q5d}). 
Note that both of the quark combinations 
$uudd\bar d$ and $uuud\bar u$ in the $\Delta^+$ contribute to the 
$qqqq\bar q\rightarrow qqq+\gamma$ transition through 
$d\bar d\rightarrow \gamma$ and $u\bar u\rightarrow \gamma$, 
respectively. This leads to the following 
factor in spin-flavor-color (SFC) space:  
\beqa
&&C_{SFC}^{3/2}=-\frac{4\sqrt{5}}{{45}}A_{p3}
(\sqrt{5}A_{\Delta 5}^{(1)}+A_{\Delta 5}^{(2)})\, , \nonumber\\ 
&&C_{SFC}^{1/2}=-\frac{8\sqrt{15}}{{135}}A_{p3}
A_{\Delta 5}^{(2)}\, .
\label{asfc}
\eeqa
Here a factor 4 standing for the number of annihilating 
quark pairs has been multiplied and the normalization 
factors of the proton wave function
in $qqq$ configuration and $\Delta$ wave function in $qqqq\bar q$ 
configuration are explicit. 
It may be seen from 
eq. (\ref{asfc}) that the $P-$shell $qqqq$ configuration in 
$\Delta$ does not 
contribute to the decay amplitudes with helicity $1/2$.  
The matrix element in orbital space may be approximately 
evaluated with a power series expansion in $k_\gamma$, with
the result \cite{delpion}:
\begin{equation}
\langle T \rangle \simeq ({\omega_3 \omega_5\over \omega^2})^3 
{k_\gamma \over \omega_5}{\sqrt{2}\over 4}
(1-{3\over 20}{k_\gamma ^2\over \omega_5^2})\,.
\label{ao}
\end{equation}
Here the normalization factor $({\omega_3 \omega_5 / \omega^2})^3$
, with $\omega=\sqrt{(\omega_3^2 +\omega_5^2)/2}$, comes from the 
different values of the size parameters $\omega_3$ in eq. (\ref{harmosc})
and $\omega_5$ in eq. (\ref{harmosc5}).

The helicity amplitudes for the direct annihilation process
are obtained by taking the product
of the matrix elements in SFC space (\ref{asfc}) and orbital 
space (\ref{ao}) as
\beqa
&& A_{a{\frac{3}{2}}}=-\frac{\sqrt{10}}{{45}}A_{p3}
(\sqrt{5}A_{\Delta 5}^{(1)}+A_{\Delta 5}^{(2)})
({\omega_3 \omega_5\over \omega^2})^3 
{e\sqrt{k_\gamma} \over \omega_5}(1-{3\over 20}
{k_\gamma ^2\over \omega_5^2})\nonumber\\
&& A_{a{\frac{1}{2}}}=-\frac{2\sqrt{30}}{{135}}A_{p3}A_{\Delta 5}^{(2)}
({\omega_3 \omega_5\over \omega^2})^3 
{e\sqrt{k_\gamma} \over \omega_5}(1-{3\over 20}
{k_\gamma ^2\over \omega_5^2})~.
\label{hela}
\eeqa
As the ratio of these amplitudes differs from $\sqrt{3}$ they
lead to a non-vanishing value for the $E2/M1$ ratio. 

Assuming again that the probability of the $qqqq\bar q$
configuration in both the proton and the $\Delta(1232)$ is
10\%, in which the $J=1$ and $J=2$ $qqqq$ states of the 
$\Delta(1232)$ are assumed to 
have the proportion of 50\% and 50\%, respectively, 
the direct annihilation transition leads to an
enhancement of the calculated value for the helicity
amplitudes $A_{{1\over 2}}$ and $A_{{3\over 2}}$ 
factors 1.01 and 1.08 respectively.
When these enhancement factors are combined with those
that arise from transitions between the $qqqq\bar q$ components,
the combined enhancement factor for the calculated value
of the helicity amplitude
$A_{{1\over 2}}$ is 1.12 and that for the amplitude
 $A_{{3\over 2}}$ is 1.18. When the standard $qqq$ quark model values,
-0.083$/\sqrt{\mathrm{GeV}}$ and
-0.143$/\sqrt{\mathrm{GeV}}$, for these two amplitudes
are multiplied by the corresponding enhancement factors, the
net calculated values for the two amplitudes become
$A_{{1\over 2}}=-0.093 /\sqrt{\mathrm{GeV}}$ and
$A_{{3\over 2}}=-0.171/\sqrt{\mathrm{GeV}}$,
respectively. The first of these
two values is smaller by a factor 1.4 than the corresponding
empirical value, and the second is smaller by a factor 1.5
than the empirical value. The calculated ratio of the
two helicity amplitudes,
$A_{{3\over 2}}/A_{{1\over 2}}=1.84$ is 
however considerably
closer to the empirical value 1.89 than the value
$\sqrt{3}\simeq 1.73$ that is obtained in the $qqq$ and 
$qqqq\bar q$ quark model. Assuming a 10\% probability of the 
$qqqq\bar q$ component in the proton and 20\% in the $\Delta(1232)$, 
with equal proportion of the $J=1$ and $J=2$ $qqqq$ states
in $\Delta(1232)$,  
the calculated helicity amplitudes increase to 
$A_{{1\over 2}}=-0.096 /\sqrt{\mathrm{GeV}}$ and
$A_{{3\over 2}}=-0.180/\sqrt{\mathrm{GeV}}$. These values
lead to the ratio
$A_{{3\over 2}}/A_{{1\over 2}}=1.88$, which very close to the 
empirical value, and to an enhancement of the calculated
decay width by 1.5. 

\subsection{Annihilation with quark-quark confinement interactions}

\begin{figure}[t]
\vspace{20pt} 
\begin{center}
\mbox{\epsfig{file=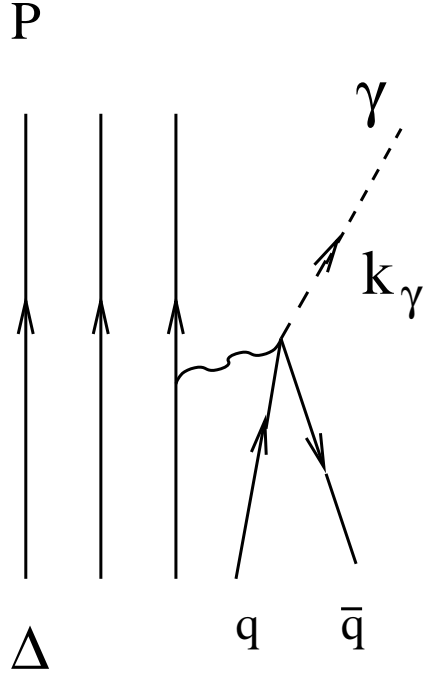, width=55mm}} 
\caption{Confinement induced
 $qqqq\bar q\rightarrow qqq\gamma$ annihilation
process}
\label{fig2}
\end{center}
\vspace{10pt} 
\end{figure}

Quark-antiquark annihilation transitions can be triggered
by the interactions between the quarks in the baryons.
The most obvious such triggering interaction is the
confining interaction.
Recently it has been noted that the confining interaction
may contribute significantly to the calculated pion 
decay width of the $\Delta$ \cite{delpion}.
A similar effect naturally also should be expected
in the case of the transition
$\Delta^+\rightarrow p\gamma$ (Fig. \ref{fig2}).
To lowest order in the quark momenta the amplitude for
this confinement triggered annihilation mechanism
may, in the case of a linear confining interaction,
be derived by making the following replacement
in the transition operator for direct annihilation (\ref{at}):
\begin{equation}
e_i{\sigma_i}_-\rightarrow e_i{\sigma_i}_- 
(1-\frac{c r_{ij}-b}{2m})\,.
\label{shift}
\end{equation}  
Here c is the string tension, $r_{ij}$ is the
distance between the two quarks that interact by the
confining interaction and b is a constant, which shifts 
the zero point of the confinement. This replacement
applies both in the case of scalar and vector
coupled confinement. 
If the confining interaction is assumed to have the
color coupling $\vec\lambda_i^C\cdot\vec\lambda_j^C$,
the string tension $c$ should be the same for all the
$qq$ and $q\bar q$ pairs in the $qqqq\bar q$ system,
and half of the value for quark pairs in 
three-quark systems \cite{richard}. Here we use the
value $c$ = 280 MeV/fm \cite{delpion}.

From this expression it follows that for $b>0$, which
implies that the effective confining potential
is negative at short range, and which is suggested
by phenomenological study of the $D$ and $D_s$ meson
spectra \cite{timoa}, the confining interaction reduces the
net annihilation amplitude. This situation is 
similar to that in electromagnetic decay of heavy
quarkonia and heavy light mesons \cite{timob,timoc}.

The orbital matrix element of the annihilation process with the linear 
interaction 
$-(cr-b)/2m$ between quarks is given by \cite{delpion}
\begin{equation}
\langle T_{conf}\rangle= -({\omega_3 \omega_5\over \omega^2})^3 
{c\over m \omega}\,({k_\gamma\over\omega_5})\,
{128\sqrt{15}\over 375\pi}\, K(k_\gamma)\, .
\label{confo}
\end{equation}
Here the function $K(q)$ is defined as
\begin{equation}
K(q)=\omega_5^5\int_0^\infty d\xi\,\xi^4\,{j_1(\beta q\xi)
\over \beta q\xi}\,e^{-\alpha^2\xi^2}\, k(\omega\xi)\,.
\label{K(xi)}
\end{equation}
The constants
$\alpha$ and $\beta$ in this expression are defined
as $\alpha=2\omega_5/\sqrt{5}$ and $\beta=2\sqrt{3}/5$, 
respectively. 
The function $k(y)$ is defined as:
\begin{equation}
k(y)=\int_0^\infty dx x^2 e^{-x^2}\int_{-1}^1 dz
\{\sqrt{x^2
-2\sqrt{2} xzy+2y^2}-{\sqrt{6}\over 2}{b\omega\over c}\}\, .
\label{k(y)}
\end{equation}
A numerical calculation gives $K(k_\gamma)$=0.71 at $
k_\gamma$=259 MeV
in the case $b=0$.

The SFC matrix element of the annihilation process with confinement is 
the same 
as that of the direct annihilation given by eq. (\ref{asfc}) and,  
combining with the orbital matrix element in eq. (\ref{confo}), one 
obtains the 
helicity amplitudes expressions:
\beqa
&& A_{a{\frac{3}{2}}}^{c}=\frac{512\sqrt{3}}
{{1125}\pi}A_{p3}(\sqrt{5}A_{\Delta 5}^{(1)}+A_{\Delta 5}^{(2)})~
({\omega_3 \omega_5\over \omega^2})^3 
{c\over m \omega}\,{e\sqrt{k_\gamma}\over\omega_5} 
K(k_\gamma)\, \nonumber\\
&& A_{a{\frac{1}{2}}}^{c}= \frac{1024}{{1125}\pi}A_{p3}A_{\Delta 5}^{(2)}~
({\omega_3 \omega_5\over \omega^2})^3 
{c\over m \omega}\,{e\sqrt{k_\gamma}\over\omega_5} K(k_\gamma)\, ,
\label{helac}
\eeqa
where an overall factor 3 has been inserted for
the 3 similar interacting processes
of annihilating the antiquark.

Due to the opposite sign of the amplitudes in eq. (\ref{helac}) 
to that in eq. 
(\ref{hel3q}), (\ref{hel5q}), (\ref{hela}), 
the confinement triggered annihilation reaction reduces the
calculated values of the helicity amplitudes, and increases
the disagreement with the empirical values.
The quantitative importance of this effect does of course
depend on the value of the constant $b$. If this is taken to
be $\sim 480$ MeV, the matrix element of the
amplitude of the confinement triggered
annihilation transition vanishes.
The contribution of
the annihilation transition triggered by linear confinement
to the $\Delta(1232)\rightarrow p\gamma$ 
decay is therefore insignificant when the constant
in the linear confining interaction is taken to
be in the range by $350-500$ MeV (Table {\ref{tab:linearamp}). 
This range is only
slightly larger than the 
values employed in the literature on heavy flavor
spectroscopy \cite{timoc}. 

\begin{table}
\caption{Calculated helicity amplitudes, the ratio between the 
helicity amplitudes and the enhancement of calculated
electromagnetic decay width from the $qqq$ quark model 
value ($\delta$) for 
different values of the constant b in the
linear confining potential. 
Here the probability
of the $qqqq\bar q$
component in both the nucleon and in the $\Delta(1232)$
is taken to be 10\%.
}
\label{tab:linearamp}
\begin{tabular}{cccccc}
\hline 

   $b$    & $K (k_\gamma)$ &    $A_{3/2}$       &    $A_{1/2}$   
  & $A_{3/2}/A_{1/2}$ &$\delta$\\
  (MeV)   &                       & $(1/\sqrt{GeV})$   & $(1/\sqrt{GeV})$
 &                   &\\
\hline

  300 & 0.26 & -0.151 & -0.086 & 1.76 &1.08\\
  350 & 0.18 & -0.156 & -0.088 & 1.78 &1.16\\
  400 & 0.11 & -0.162 & -0.090 & 1.80 &1.23\\
  450 & 0.03   & -0.167 & -0.092 & 1.82 &1.31\\
  500 & -0.04& -0.173 & -0.094 & 1.84 &1.39\\
  550 & -0.12& -0.178 & -0.096 & 1.86 &1.48\\

\hline
\end{tabular}
\end{table}

To have an estimate of the theoretical
uncertainty of the magnitude of the
contribution of the confinement triggered
annihilation process, we also consider the case of
harmonic confinement, which is consistent with the
wave function model.
This is obtained by the substitution \cite{delpion}:
\begin{equation}
cr-b\rightarrow {1\over 2}C r^2-B\, .
\label{osc}
\end{equation}
Here $B$ is a constant that shifts the interaction
potential to negative values at short range.
The oscillator constant $C$ is given as \cite{helminen}
\begin{equation}
C = {m\omega_5^2\over 5}\, .
\label{bigC}
\end{equation}
With $m=340$ MeV and $\omega_5=245$ MeV this gives for
$C$ the value 105 MeV/fm$^2$.

\begin{table}
\caption{Calculated helicity amplitudes, the ratio between the 
helicity amplitudes and the enhancement of calculated
electromagnetic decay width from the $qqq$ quark model 
value ($\delta$) for 
different values of the constant B in the
harmonic confining potential.
Here the probability
of the $qqqq\bar q$
component in both the nucleon and in the $\Delta(1232)$
is taken to be 10\%.}
\label{tab:oscamp}
\begin{tabular}{cccccc}
\hline 
   $B$    & $L (k_\gamma)$ &    $A_{3/2}$       &    $A_{1/2}$   
  & $A_{3/2}/A_{1/2}$ &$\delta$\\
  (MeV)   &                       & $(1/\sqrt{GeV})$   & $(1/\sqrt{GeV})$
 &                   &\\
\hline

  50  & 1.41 & -0.157 & -0.088 & 1.78&1.17\\
  100 & 0.82 & -0.162 & -0.090 & 1.80&1.24\\
  150 & 0.23   & -0.168 & -0.092 & 1.83&1.32\\
  200 & -0.36&-0.173 & -0.094 & 1.85&1.40\\
  250 & -0.95& -0.179& -0.096 & 1.87&1.49\\
  300 & -1.54& -0.184 & -0.098 & 1.89&1.57\\

\hline
\end{tabular}
\end{table}

The helicity amplitude for confinement triggered annihilation
$\Delta^{+}\rightarrow p\gamma$ in this oscillator
confinement model  may be obtained directly from
the expression for linear confinement above
(\ref{helac}) by the substitution \cite{delpion}
\begin{equation}
c K(k_\gamma)\,\rightarrow \, {\sqrt{6}\over 6}{C\over \omega}
L(k_\gamma)\,.
\end{equation}
The function $L(q)$ is defined as the integral
\begin{equation}
L(q) = \sqrt{\pi}\omega_5^5\, \int_0^\infty d\xi\,
\xi^4\,{j_1(\beta  q \xi)\over \beta q\xi}\,
e^{-\alpha^2\xi^2}\,({3\over 4}+\omega^2
(\xi^2-{3B\over 2C}))\,.
\label{L(xi)}
\end{equation}
For $\Delta^+ (1232)
\rightarrow p\gamma$\,, $k_\gamma$=259 MeV and 
$L(k_\gamma)$=2.0 in the case where $B=0$. 

The confinement triggered annihilation transitions also in this
case counteracts the contribution from the direct annihilation
transition, unless $B$ takes a very large value. The contribution
from the confinement triggered annihilation to
the helicity 
amplitudes of the $\Delta^{+}\rightarrow p\gamma$ decay
vanishes if $B$ takes the value $B\approx 170$ MeV.
For values of $B$ in the range $100-200$ MeV, 
the confinement triggered
annihilation transitions are insignificant 
(Table {\ref{tab:oscamp}).

\section{Discussion}

The results obtained above with an extension of the $qqq$
quark model to include $10-20$\% admixtures of the $qqqq\bar q$
configurations, that are expected to have the lowest
energy, reveal that direct annihilation transitions 
of the form $qqqq\bar q\rightarrow qqq\gamma $ 
significantly reduce
the difference between the calculated and the empirical
values of the helicity
amplitudes for $\Delta\rightarrow N\gamma$ decay.
As in addition the increase of the calculated
$A_{3\over 2}$ helicity amplitude is larger than that
of the $A_{1\over 2}$ amplitude, the calculated 
ratio $A_{3\over 2}/A_{1\over 2}$ may be brought into agreement
with the empirical values by introduction
of such $qqqq\bar q$ admixtures into the quark model
wave functions for the nucleon and the $\Delta(1232)$
resonance.

These results are consistent with the conclusion based
on studies of the coupled channel $N-\Delta-\pi$
system, that the effect of the ``pion cloud'' around
the baryons is significant, and responsible for $\sim$
30\% of the $N-\Delta$ transition magnetic moment. 

The change of the calculated ratio of the helicity
amplitudes $A_{3\over 2}/A_{1\over 2}$ from the quark model
value $\sqrt{3}$ to 1.88 is consistent with the
magnitude expected on the basis of the large $N_C$
limit of QCD \cite{jixd}: 
\begin{equation}
A_{3\over 2}/A_{1\over 2} = \sqrt{3} + O(1/N_c^2)\, .
\end{equation}
Equivalently, 
the ratio $E_2/M_1$ of the multipole amplitudes is predicted to be 
order $1/N_c^2$.

Introduction of $10-20$\% admixtures of $qqqq\bar q$
components in the wave functions of the nucleon and the
$\Delta(1232)$ resonance was however not found to be
sufficient to completely remove the underprediction of
the empirical values of the helicity amplitudes
$A_{1\over 2}$ and  $A_{3\over 2}$ in the
quark model. Additional annihilation mechanisms that
are triggered by the interaction between the
quarks may be required for this purpose. As an
example of such a mechanism the annihilation
mechanism that is triggered by the confining
interaction was considered here. In the case of
linear confinement it was however found that this
mechanism only leads to an additional enhancement
if the linear interaction potential is large
an negative at short range, which may not be
phenomenologically realistic.

\begin{acknowledgments}

Research supported in part by the Academy of Finland grant 
number 54038 

\end{acknowledgments}

\end{document}